# BreastRegNet: A Deep Learning Framework for Registration of Breast Faxitron and Histopathology Images*

Negar Golestani[1], Aihui Wang[2], Gregory R Bean[2], and Mirabela Rusu[1,3,4]

[1] Department of Radiology, Stanford University, CA, USA
[2] Department of Pathology, Stanford University, CA, USA
[3] Department of Urology, Stanford University, CA, USA
[4] Department of Biomedical Data Science, Stanford University, CA, USA
{negaar, mirabela.rusu}@stanford.edu

**Abstract.** A standard treatment protocol for breast cancer entails administering neoadjuvant therapy followed by surgical removal of the tumor and surrounding tissue. Pathologists typically rely on cabinet X-ray radiographs, known as Faxitron, to examine the excised breast tissue and diagnose the extent of residual disease. However, accurately determining the location, size, and focality of residual cancer can be challenging, and incorrect assessments can lead to clinical consequences. The utilization of automated methods can improve the histopathology process, allowing pathologists to choose regions for sampling more effectively and precisely. Despite the recognized necessity, there are currently no such methods available. Training such automated detection models require accurate ground truth labels on ex-vivo radiology images, which can be acquired through registering Faxitron and histopathology images and mapping the extent of cancer from histopathology to x-ray images. This study introduces a deep learning-based image registration approach trained on mono-modal synthetic image pairs. The models were trained using data from 50 women who received neoadjuvant chemotherapy and underwent surgery. The results demonstrate that our method is faster and yields significantly lower average landmark error (2.1±1.96 mm) over the state-of-the-art iterative (4.43 ± 4.1 mm) and deep learning (4.02 ± 3.15 mm) approaches. Improved performance of our approach in integrating radiology and pathology information facilitates generating large datasets, which allows training models for more accurate breast cancer detection.

## 1 Introduction

Breast cancer is a prevalent and fatal disease, and it ranks as the most frequently diagnosed cancer among women in several nations, including the United States

---

* Supported by the Department of Radiology at Stanford University, Philips Healthcare, Stanford Cancer Imaging Training Program (T32 CA009695), and National Cancer Institute (R37CA260346). The content is solely the responsibility of the authors and does not necessarily represent the official views of the National Institutes of Health.

[1]. The diagnostic process usually involves mammography, breast ultrasound, and biopsy, with treatment options dependent on diagnosis results. One standard treatment approach for breast cancers meeting certain clinicopathologic criteria involves neoadjuvant therapy followed by surgery [2]. After surgery, pathologists analyze excised tissue specimens to obtain information on tumor size, grade, stage, and margin status of residual cancer, which is crucial in determining further treatment options. Pathology processing includes imaging gross sections using cabinet x-ray radiographs (Faxitron), with only a select few representative sections being further processed for histology. Pathologists manually select these representative samples based on gross examination and Faxitron radiographs, but it is an estimation and subject to error.

Accurate identification of the tumor site and extent of the disease is a challenging task that can lead to delays in the pathology process and require additional follow-up if the initial estimation is inaccurate [3]. The automated identification of residual tumors or tumor bed on excised tissue radiographs can significantly improve the pathology workflow and diagnostic accuracy, ultimately leading to faster turnaround for results and improved prognosticating of patient outcomes. However, no automated methods currently exist to differentiate residual tumors from reactive stromal changes on Faxitron radiographs. To address this limitation, the registration of Faxitron and histopathology images represents a critical step towards accurately mapping the extent of cancer from histopathology images onto their corresponding Faxitron radiographs of ex-vivo tissue. However, aligning the two images poses three major challenges: variations in cancer appearance across different modalities, imprecise correspondence due to a rough estimate of sampled tissue within the Faxitron image, and differences in image format and content, with Faxitron images displaying a projection of the entire gross slice or macrosection in the x-ray, whereas pathology images are $5\mu m$ sections at various depths through the tissue block, leading to potential inaccuracies in registration.

Multi-modal image registration can be achieved using traditional iterative or deep learning approaches. Iterative methods minimize a cost function through optimization techniques, but they can be computationally demanding and may easily be trapped in local optima, resulting in incorrect alignments [4–8]. In contrast, deep learning-based image registration trains neural networks to align moving images with fixed images, which eliminates the need for an iterative optimization process and can directly align input images, thereby speeding up the registration process [9–16]. Although deep learning has been successful in medical image registration, many studies have relied on supervised approaches that require large labeled datasets [17,18]. However, our Faxitron-Histopathology dataset lacks ground-truth registered images, making it unsuitable for supervised methods. Unsupervised techniques can address the limited training data issue, but they mainly focus on mono-modal registration, with limited studies on multimodal registration [17].

We introduce the breast registration network (BreastRegNet), a deep learning approach for affine registration of breast Faxitron and histopathology images.

To avoid the need for ground truth alignment, we employ weakly supervised strategies during training and integrate unsupervised intensity, segmentation, and regularization terms in our loss function. We train the network on synthetic mono-modal data and their tissue segmentation masks to overcome the limitations of multi-modal similarity measures, which are unsuitable for this problem due to the uncorrelated intensities of Faxitron and histopathology images. During training, the standard distribution distance measure incorporated as the regularizer loss term trains the network to extract input image representations with minimized distribution distance. It enables the network to accurately process multi-modal data at the inference stage without requiring tissue segmentations, as it has learned to solve image registration problems regardless of the image modalities. The main contributions of our study are summarized as follows:

- Introduction of the first deep-learning approach designed to register breast Faxitron-histopathology images.
- Implementation of a weakly supervised network that tackles two main challenges in registering multi-modal data: absence of ground-truth training data and limitations of multi-modal similarity metrics.
- Utilization of domain confusion loss as the regularization term to enable joint optimization of image features for registration and domain invariance.

## 2   Methods

### 2.1   Data Description and Analysis

This study, approved by the institutional review board, includes data from 50 women participants who received neoadjuvant chemotherapy and subsequently underwent surgical excision. Tissue samples were processed according to standard protocols, with macrosection distances varying based on the specimen size (∼3mm for lumpectomies and up to ∼1cm for mastectomies). A Faxitron radiograph with a pixel size of 3440 × 3440 was obtained for each patient to provide an ex-vivo view of the macrosections of resected and sectioned tissue. Depending on the specimen size, histological examination was conducted on either the entire excision or specific sections. Therefore, digital hematoxylin and eosin (H&E) images with a pixel size of 4600 × 6000 were acquired for the tissue segments with their approximate location annotated as labeled box regions of interest (Box-ROIs) on the corresponding Faxitron image by the pathologists.

The histopathology and Faxitron images corresponding to each tissue segment were extracted and processed. The resulting dataset consisted of 1093 pairs of histopathology and Faxitron images. A breast subspecialty pathologist evaluated the histopathology images using clinical reports and re-reviewed all cases to annotate the extent of residual cancer or tumor bed in cases of complete pathologic response. Tissue masks on both images were automatically extracted and manually verified. Furthermore, matching landmarks were manually annotated based on visual similarities by an expert image analyst with two years of experience reviewing Faxitron and histopathology images of the breast.

## 2.2   Registration Network

Our proposed network for registering Faxitron and histopathology images of the breast is shown in Fig. 1. In this work, we considered the Faxitron and histopathology images as the fixed and moving images, respectively, and resampled them to 224 × 224 pixels before feeding them into the network. The registration network consists of two separate feature extraction networks, each tailored for one of the image types (i.e., Faxitron or histopathology), to capture the discriminative features of the input images. We employed pre-trained VGG-16 networks, cropped at the fourth block layer, followed by per-feature L2 normalization. These networks, trained on the ImageNet dataset [19], take the moving image $I_m \in R^{H \times W \times 3}$ and the fixed image $I_f \in R^{H \times W \times 3}$ as inputs to produce their corresponding feature maps $F_m, F_f \in R^{H \times W \times d}$. The resulting feature images, with a grid size of $H \times W$ and d-dimensional voxel vectors, are subsequently normalized and provided as input to a correlation layer.

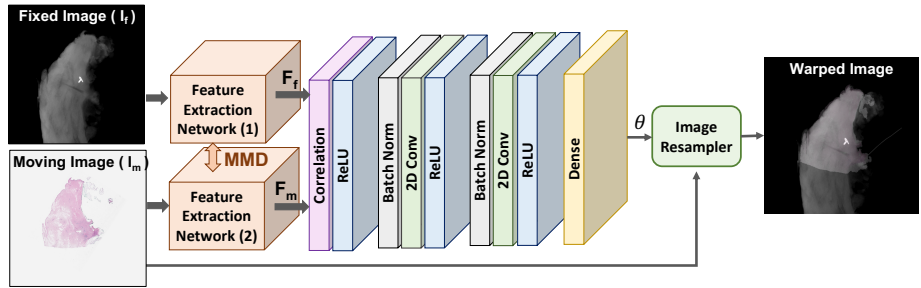

Fig.1: Overview of the proposed deep learning-based registration network.

This correlation layer integrates the feature images into a single correlation map $C_{mf}$ of the same size and voxel length of $H \times W$, containing all pairwise correlation coefficients between feature vectors [20], as given by:

$$C_{mf}(i.j,k) = \frac{\text{cov}\left[F_m(i,j) \cdot F_f(p,q)\right]}{\text{std}\left[F_m(i,j)\right] \text{std}\left[F_f(p,q)\right]} \quad (1)$$

where $k = p + H(q - 1)$, and $F_m(i,j)$ and $F_f(p,q)$ are the feature vectors positioned at $(i,j)$ and $(p,q)$ in the feature maps corresponding to moving and fixed images, respectively. The notation cov[.] indicates covariance, and std[.] denotes standard deviation.

The correlation map obtained is subjected to normalization by applying a rectified linear unit (ReLU) and channel-wise L2 normalization at each spatial location to down-weight ambiguous matches. The normalized correlation map $F_{mf}$ is then passed through two stacked blocks of convolutional layers and a fully connected layer. Each convolutional block consists of a convolutional unit, batch normalization, and ReLU activation designed to reduce the dimensionality of the correlation map. The network outputs a vector of length six representing the affine transformation parameters between the input images.

Instead of using direct estimations as the transformation matrix, we employed the approach proposed in [14], which adds an identity transform to scaled parameters. This technique keeps the initial estimation close to the identity map, thereby improving the stability of the model. The final matrix is defined as $\theta = \alpha\hat{\theta} + \theta_{Id}^{aff}$, where $\hat{\theta}$ is the estimated affine matrix by the network, $\theta_{Id}^{aff}$ is the parameter vector for affine identity transform, and the scaling factor $\alpha$ was set to a small constant of 0.1. Using the final transformation matrix, original histopathology images, cancer labels, and landmarks can be mapped to their corresponding Faxitron images. It should be noted that the registration network estimates the transformation parameters using preprocessed and resized Faxitron and histopathology images, while the resampler takes original high-resolution histopathology images and their corresponding data to produce the warped sample.

## 2.3 Training

Due to the absence of ground truth spatial correspondences between histopathology images and their corresponding Faxitron images, we employed synthetic transformations to generate mono-modal image pairs for training our neural networks. The parameters of the affine transformations were randomly sampled with constraints such as rotation angle between -20 to 20 degrees, scaling coefficients between 0.9 to 1.1, translation coefficients within 20% of the image size, and shearing coefficients within 5%. We created a random transformation matrix for each histopathology and Faxitron image and used the deformed images as training samples, along with their original image and transformation matrix. The network is trained by minimizing a loss function defined as:

$$L = L_{int} + L_{seg} + \lambda L_{reg} \qquad (2)$$

where $L_{int}$, $L_{seg}$, and $L_{ref}$ denote image similarity loss, segmentation loss, and regularization loss, respectively. The hyperparameter $\lambda$ = 0.01 controls the contribution of the regularization objective.

We employed the mean squared error (MSE) as an unsupervised measure of intensity loss between the fixed image $I_f$ and the warped images $I_w$, as the training dataset consisted of synthetic mono-modal samples. Furthermore, we defined the segmentation loss $L_{seg}$ as the Dice loss between the fixed tissue mask $M_f$ and the warped tissue mask $M_w$. During training, the inclusion of this loss term allows models to learn from tissue masks and eliminates the necessity for masks during inference, resulting in a weakly-supervised approach.

Training a network to estimate transformation parameters using only synthetic mono-modal samples can result in overfitting the model in processing images from the same modality, leading to a suboptimal performance on test data involving multi-modal samples. To address this issue, inspired by domain adaptation approaches [21,22], we utilized the maximum mean discrepancy (MMD) as a regularization loss function to measure the distance between the distributions of fixed and moving features produced by two feature extraction networks. By minimizing the discrepancy between feature representations, the two networks can be trained to generate feature representations that are more similar to each other. It allows us to leverage the neural network trained on mono-modal samples during the training phase to estimate the transformation parameters at inference time when the input samples originate from different modalities and feature representations from distinct domains.

### 2.4    Evaluation Metrics

We used two metrics to evaluate the alignment accuracy between the moving (histopathology) and corresponding fixed (Faxitron) images. The first metric was the total execution time (ET), which represents the time required for the approach to execute. The second metric used was the mean landmark error (MLE) defined as:

$$MLE = \frac{1}{N} \sum_{i=1}^{N} \| p_i - \phi_\theta(q_i) \| \tag{3}$$

where $\{p_i\}_{i=1}^{N}$ and $\{q_i\}_{i=1}^{N}$ denote the *N* pairs of landmarks in fixed and moving images. The parameter $\phi_\theta$ represents a transformation parameterized by $\theta$.

### 2.5    Implementation Details

The models were implemented using PyTorch [23] and trained on an NVIDIA RTX A6000 GPU and an Intel Core i9-10900K CPU (with 16GB of memory and a 3.70 GHz clock speed). We employed the Adam optimization algorithm [24] with an initial learning rate of $10^{-4}$, a learning rate decay of 0.95, a step size of 1, and a batch size of 64 to train the networks. We used 5-fold cross-validation to partition the patient data. For each fold, we trained the model on synthetic data generated from the data of four folds over 50 epochs and then tested it on the clinical data of patients within the remaining fold. The code is available online at: https://github.com/pimed/BreastRegNet.

## 3    Results

We evaluated the performance of our deep learning model in comparison to existing multi-modal registration techniques, including an iterative approach and a deep learning-based method. The iterative registration method available from SimpleITK [5,6] was implemented as a configurable multi-resolution registration approach trained using normalized mutual information (NMI) loss. Deep learning-based approaches were also employed as baselines for comparative analysis. We utilized the CNNGeometric approach [9], a deep learning network trained on a synthetic dataset with generated affine transformations. The training employed a loss function based on point location differences. Additionally, we implemented ProsRegNet [14], a CNN-based model for MRI-histopathology image registration in prostate cancer. Another baseline approach, C2FViT [25], a coarse-to-fine vision transformer model for affine transformations, was also included.

To ensure consistency and fairness in the evaluation, all models in our study were trained under similar conditions as our proposed BreastRegNet method. The hyperparameters used in each model were based on their original studies. Furthermore, as presented in our model, we parametrized the affine transformations using a weighted sum of an identity transform and the estimated parameter vector for improved stability and robustness.

Fig. 2 illustrates the outcomes acquired for the upper and lower regions of a breast tissue segment, corresponding to the superior and inferior halves. The alignment examples generated by our BreastRegNet indicate that the boundaries of breast tissue between the Faxitron and histopathology sections are precisely aligned, demonstrating the efficacy of the method in achieving an accurate global alignment of the tissue. These results suggest that, despite being trained on mono-modal input, our model can process multi-modal data and improve alignment accuracy, as evidenced by the reduction in landmark error. Moreover, the outcomes suggest that iterative registration methods, such as those provided by SimpleITK, may exhibit suboptimal performance in specific scenarios, even inferior to not performing registration. One possible explanation is the inability of the algorithm to handle missing data effectively. For example, when there is a significant mismatch in tissue areas between Faxitron and histopathology images, the SimpleITK iterative registration may attempt to align the histopathology image with the complete tissue in the Faxitron segment without considering shape matching at the boundaries. Such issues can result in poor performance of the method in the alignment of Faxitron and histopathology images.

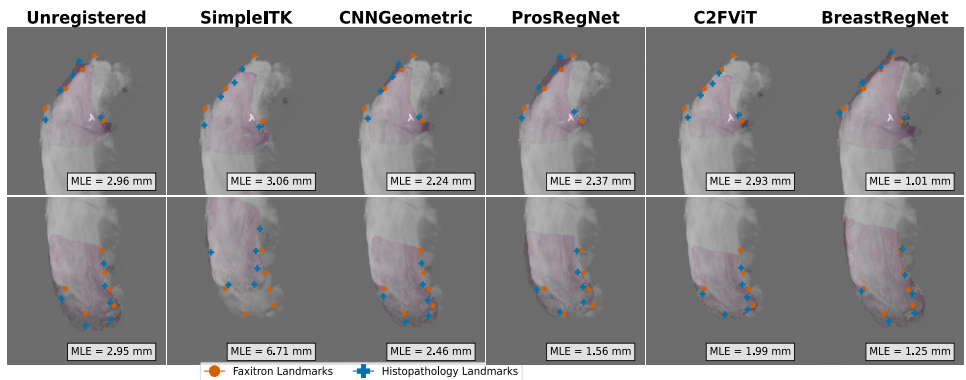

Fig.2: Overlay of transparent histology onto Faxitron pre-registration and post-registration of two segments with their corresponding landmarks.

Figure 3 presents the MLE results of our registration model compared to other approaches. The findings demonstrate that BreastRegNet outperformed both iterative and deep learning registration techniques with statistical significance. Moreover, the model performs better when trained in a weakly-supervised manner that includes segmentation loss compared to its unsupervised counterpart (US-BreastRegNet). We employed the Mann-Whitney test to evaluate the statistical significance of the comparison between registration models. Furthermore, the average execution time for the BreastRegNet was 0.07 seconds, while the iterative SimpleITK model, CNNGeometric, ProsRegNet, and C2FViT had a running time of 10.3, 0.32, 0.55, and 0.12 seconds. In conclusion, our proposed deep learning approach offers improved alignment of tissue boundaries while being faster than other registration approaches.

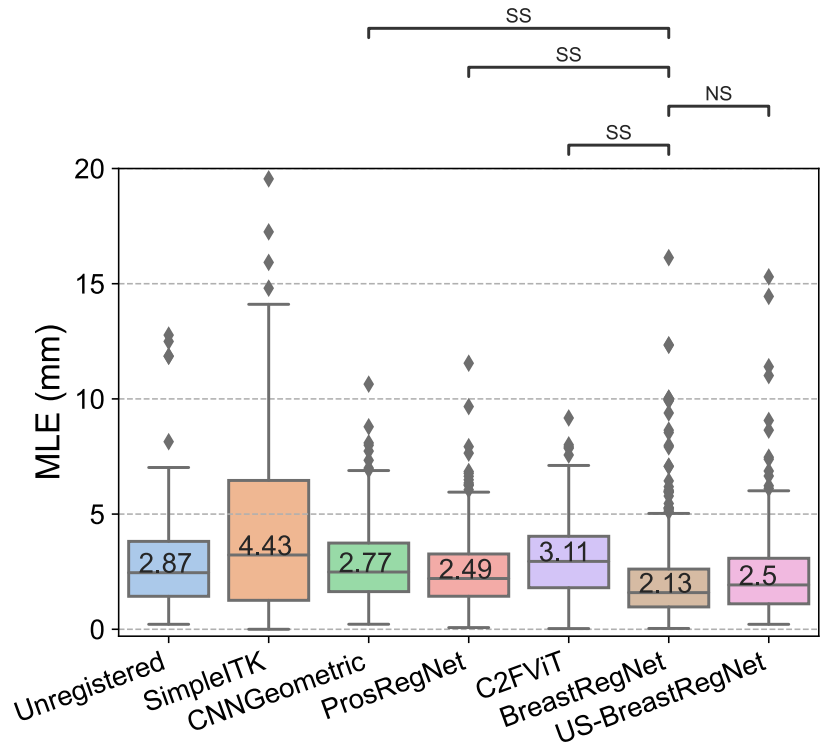

*Fig. 3: Mean landmark error comparison before and after registration.*
SS: statistically significant (p≤0.01); NS: not significant.

**Ablation Study** In our experiments, different pre-trained deep neural networks were utilized as feature extraction networks of the proposed registration model. Specifically, ResNet101 [26], VGG16 trained on the ImageNet dataset [19], and ResNet50 trained on the open-access medical image database RadImageNet (RadResNet50) [27] were employed. The performance of these models was assessed under two scenarios where the layers were either frozen or the final layer was made trainable. The VGG16 model trained on ImageNet with a trainable layer demonstrated superior performance among all these deep learning-based networks. In order to assess the impact of regularization loss, we employed $\lambda$ as a weight factor with a value of zero resulting in training the model without domain adaptation. We conducted experiments using different values of $\lambda$ and determined that a value of 0.01 yielded the best MLE performance. Comparative results are presented in Fig. 4.

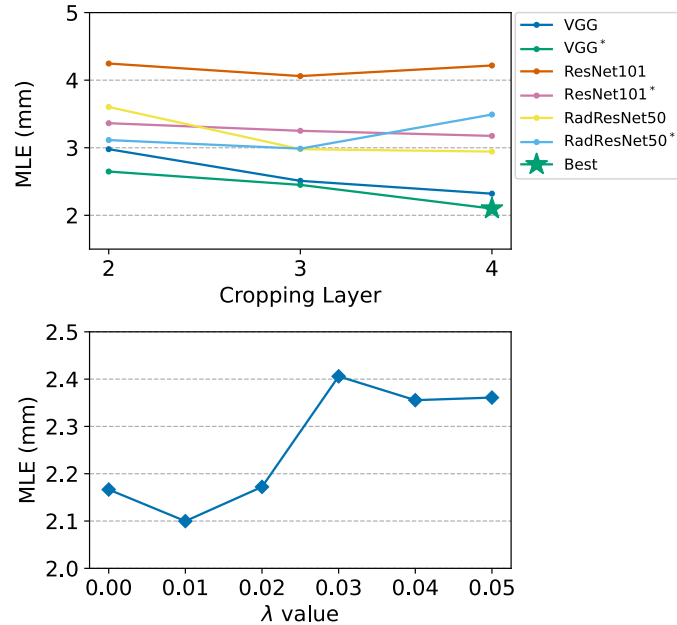

Fig.4: Performance comparison of feature extraction networks cropped at different layers (top) and different $\lambda$ values (bottom).
* denotes fine-tuned network.

## 4  Discussion and Conclusion

This paper presents a weakly supervised learning method for the registration of Faxitron and histopathology images of the breast in patients who underwent surgery. The study compares the performance of several methods and demonstrates that the proposed deep learning approach surpasses the existing multimodal registration method, including traditional iterative and deep learning techniques. This pipeline substantially improves landmark error and execution time, overcoming the limitations of conventional methods that are sensitive to initialization parameters and have lengthy computation times. Additionally, our method does not necessitate manual segmentation of tissue during inference. Its rapid image alignment capability makes it a valuable tool for registering Faxitron and histopathology images and producing ground truth data to facilitate the development of models that aid pathologists in tissue processing. In future research, efforts will be made to explore other architectures and test the model on a larger dataset with additional cohorts and multi-reader landmarks to enhance the registration of breast Faxitron and histopathology images, as well as assess its robustness, generalizability, and independence from specific input cohorts.

## References

1. American Cancer Society. American cancer society. Eprint https://www.cancer.org/cancer/breast-cancer/about.html, 2023. Accessed: 2023-02-07.
2. Kara L Britt, Jack Cuzick, and Kelly-Anne Phillips. Key steps for effective breast cancer prevention. *Nature Reviews Cancer*, 20(8):417–436, 2020.
3. Susan C Lester. *Manual of Surgical Pathology: Expert Consult-Online and Print*. Elsevier Health Sciences, 2010.
4. Stefan Klein, Marius Staring, Keelin Murphy, Max A Viergever, and Josien PW Pluim. Elastix: a toolbox for intensity-based medical image registration. *IEEE transactions on medical imaging*, 29(1):196–205, 2009.
5. Bradley C Lowekamp, David T Chen, Luis Ibañez, and Daniel Blezek. The design of simpleitk. *Frontiers in neuroinformatics*, 7:45, 2013.
6. Ziv Yaniv, Bradley C Lowekamp, Hans J Johnson, and Richard Beare. Simpleitk image-analysis notebooks: a collaborative environment for education and reproducible research. *Journal of digital imaging*, 31(3):290–303, 2018.
7. Josien PW Pluim, JB Antoine Maintz, and Max A Viergever. Mutual-informationbased registration of medical images: a survey. *IEEE transactions on medical imaging*, 22(8):986–1004, 2003.
8. Dinggang Shen and Christos Davatzikos. Hammer: hierarchical attribute matching mechanism for elastic registration. *IEEE transactions on medical imaging*, 21(11):1421–1439, 2002.
9. Ignacio Rocco, Relja Arandjelovic, and Josef Sivic. Convolutional neural network architecture for geometric matching. In *Proceedings of the IEEE conference on computer vision and pattern recognition*, pages 6148–6157, 2017.
10. Guha Balakrishnan, Amy Zhao, Mert R Sabuncu, John Guttag, and Adrian V Dalca. Voxelmorph: a learning framework for deformable medical image registration. *IEEE transactions on medical imaging*, 38(8):1788–1800, 2019.
11. Bob D De Vos, Floris F Berendsen, Max A Viergever, Hessam Sokooti, Marius Staring, and Ivana Išgum. A deep learning framework for unsupervised affine and deformable image registration. *Medical image analysis*, 52:128–143, 2019.
12. Xingyu Jiang, Jiayi Ma, Guobao Xiao, Zhenfeng Shao, and Xiaojie Guo. A review of multimodal image matching: Methods and applications. *Information Fusion*, 73:22–71, 2021.
13. Anil Rahate, Rahee Walambe, Sheela Ramanna, and Ketan Kotecha. Multimodal co-learning: challenges, applications with datasets, recent advances and future directions. *Information Fusion*, 81:203–239, 2022.

14. Wei Shao, Linda Banh, Christian A Kunder, Richard E Fan, Simon JC Soerensen, Jeffrey B Wang, Nikola C Teslovich, Nikhil Madhuripan, Anugayathri Jawahar, Pejman Ghanouni, et al. Prosregnet: A deep learning framework for registration of mri and histopathology images of the prostate. *Medical image analysis*, 68:101919, 2021.
15. Tongxue Zhou, Su Ruan, and Stéphane Canu. A review: Deep learning for medical image segmentation using multi-modality fusion. *Array*, 3:100004, 2019.
16. Theodoros Georgiou, Yu Liu, Wei Chen, and Michael Lew. A survey of traditional and deep learning-based feature descriptors for high dimensional data in computer vision. *International Journal of Multimedia Information Retrieval*, 9(3):135–170, 2020.
17. Yabo Fu, Yang Lei, Tonghe Wang, Walter J Curran, Tian Liu, and Xiaofeng Yang. Deep learning in medical image registration: a review. *Physics in Medicine & Biology*, 65(20):20TR01, 2020.
18. Grant Haskins, Uwe Kruger, and Pingkun Yan. Deep learning in medical image registration: a survey. *Machine Vision and Applications*, 31(1):1–18, 2020.
19. Jia Deng, Wei Dong, Richard Socher, Li-Jia Li, Kai Li, and Li Fei-Fei. Imagenet: A large-scale hierarchical image database. In *2009 IEEE conference on computer vision and pattern recognition*, pages 248–255. Ieee, 2009.
20. K Kavitha and B Thirumala Rao. Evaluation of distance measures for feature based image registration using alexnet. *arXiv preprint arXiv:1907.12921*, 2019.
21. Eric Tzeng, Judy Hoffman, Ning Zhang, Kate Saenko, and Trevor Darrell. Deep domain confusion: Maximizing for domain invariance. *arXiv preprint arXiv:1412.3474*, 2014.
22. Mei Wang and Weihong Deng. Deep visual domain adaptation: A survey. *Neurocomputing*, 312:135–153, 2018.
23. Adam Paszke, Sam Gross, Francisco Massa, Adam Lerer, James Bradbury, Gregory Chanan, Trevor Killeen, Zeming Lin, Natalia Gimelshein, Luca Antiga, et al. Pytorch: An imperative style, high-performance deep learning library. *Advances in neural information processing systems*, 32, 2019.
24. Diederik P Kingma and Jimmy Ba. Adam: A method for stochastic optimization. *arXiv preprint arXiv:1412.6980*, 2014.
25. Tony CW Mok and Albert Chung. Affine medical image registration with coarse-to-fine vision transformer. In *Proceedings of the IEEE/CVF Conference on Computer Vision and Pattern Recognition*, pages 20835–20844, 2022.
26. Kaiming He, Xiangyu Zhang, Shaoqing Ren, and Jian Sun. Deep residual learning for image recognition. In *Proceedings of the IEEE conference on computer vision and pattern recognition*, pages 770–778, 2016.
27. Xueyan Mei, Zelong Liu, Philip M Robson, Brett Marinelli, Mingqian Huang, Amish Doshi, Adam Jacobi, Chendi Cao, Katherine E Link, Thomas Yang, et al. Radimagenet: An open radiologic deep learning research dataset for effective transfer learning. *Radiology: Artificial Intelligence*, 4(5):e210315, 2022.